\documentclass[fleqn,twoside,twocolumn,nofootinbib,showkeys]{revtex4} 
\usepackage[sec,nocpr]{ujp} 

\begin{document}
\title[T-Matrix in Discrete Oscillator Representation]
{T-MATRIX IN DISCRETE\\ OSCILLATOR REPRESENTATION}
\author{V.S.~Vasilevsky}
\affiliation{Bogolyubov Institute for Theoretical Physics, Nat. Acad. of Sci. of Ukraine}%
\address{14b, Metrolohichna Str., Kyiv 03680, Ukraine}%
\email{VSVasilevsky@gmail.com}
\author{M.D.~Soloha-Klymchak}%
\affiliation{Bogolyubov Institute for Theoretical Physics, Nat. Acad. of Sci. of Ukraine}%
\address{14b, Metrolohichna Str., Kyiv 03680, Ukraine}%
\udk{539.1} \pacs{03.65.-w,\\[-3pt] 03.65.Nk, 24} \razd{\secii}

\autorcol{V.S.\hspace*{0.7mm}Vasilevsky,
M.D.\hspace*{0.7mm}Soloha-Klymchak}

\setcounter{page}{297}%

\begin{abstract}
We investigate T-matrix for bound and continuous-spectrum states in
the discrete oscillator representation.\,\,The investigation is
carried out for a model problem~-- the particle in the field of a
central potential.\,\,A system of linear equations is derived to
determine the coefficients of the T-matrix expansion in the
oscillator functions.\,\,We selected four potentials (Gaussian,
exponential, Yukawa, and square-well ones) to demonstrate
peculiarities of the T-matrix and its dependence on the potential
shape.\,\,We also study how the T-matrix expansion coefficients
depend on the parameters of the oscillator basis such as the
oscillator length and the number of basis functions involved in
calculations.
\end{abstract}
\maketitle

\section{Introduction}

We are going to consider the convergence of a wave function and the
T-matrix expansion in the oscillator functions.\,\,This
consideration will be restricted to a model problem of the particle
in the field of a central potential.\,\,The analysis will be done
within the matrix form of quantum mechanics, which involves an
infinite set of oscillator functions to realize a discrete
representation.\,\,This matrix form is well known as the algebraic
version of the resonating group method or J-matrix method.\,\,The
methods were formulated in \cite{kn:Fil_Okhr, Filippov:1980mn} and
\cite{kn:Heller1, kn:Yamani}, respectively.\,\,Now, they are widely
used to describe nuclear, atomic, and molecular systems.\,\,In Ref.
\cite{J-matrix2008B}, one can find progress in resolving the
internal problems of the method and the numerous applications to
solving real physical problems in different branches of quantum
physics.

Usually, the expansion of a wave function in the oscillator or any
other square-integrable basis is considered within the
J-matrix.\,\,The discrete form of the T-matrix  has not been
investigated yet.\,\,For instance, in Ref.
\cite{2001PAN....64.1799R}, the oscillator basis was used to
construct wave functions and to calculate the phase shift for a
Gaussian potential.\,\,The T-matrix was also obtained in that paper,
but only in the momentum space.\,\,So, we are going to fill in this
gap.\,\,We consider the states of both discrete and continuous
spectra.\,\,However, the main attention will be paid to the
scattering states.

We are going to demonstrate the convergence of the T-matrix and how
the convergence depends on the shape of a potential, on the
oscillator length of a basis and the energy of the state.\,\,For
this aim, we selected four potentials, which mimic different
physical systems.\,\,Note that, by solving model problems, one can
reveal the interesting features and peculiarities of systems under
consideration, which are observed in more complicated and realistic
systems.\,\,For instance, it was discovered in Refs.
\cite{kn:AM_AJP, kn:VA_PR}, while studying simple model problems,
that the J-matrix in some cases suffers of a slow
convergence.\,\,This means that one needs to involve a very large
basis of oscillator functions to achieve the desired precision of
calculations.\,\,An effective method was formulated in
\cite{kn:VA_PR, 2004nucl.th..12085B, J-matrix2008.117}.\,\,It allows
one to reduce the set of oscillator functions by three-five times in
order to obtain the phase shifts with higher precision.

The paper is organized in the following way.\,\,In Section
\ref{sec:Method}, we make all necessary definitions and deduce the
equations for the T-matrix expansion coefficients.\,\,We briefly
consider main equations, which are used to describe quantum
systems.\,\,These equations will be transformed from a continuous
(coordinate or momentum) representation to a discrete, oscillator
representation.\,\,The analysis of the T-matrix in the discrete
representation is presented in Section \ref{sec:Analysis} for four
potentials (Gaussian, Yukawa, exponential, and square-well ones).

\section{Model Formulation}

\label{sec:Method}

We make use of the system of units by selecting the constant
$\hbar^{2}/m=1$.\,\,This leads to a renormalization of the potential
energy operator
\[
\widehat{V}\left(  r\right)
\Rightarrow\frac{m}{\hbar^{2}}\widehat{V}\left( r\right)\!.
\]
In this representation, the kinetic energy operator is $\widehat{T}=-\frac{1}%
{2}\mathbf{\nabla}^{2}$ in coordinate space and $\widehat{T}=\frac{1}%
{2}\mathbf{p}^{2}$ in the momentum space, and the energy is
$E=\frac{1}{2}k^{2}$.\,\,In this paper, we will consider central
potentials.\,\,Thus, the orbital momentum $L$ is a good quantum
number.

\subsection{Basic equations}

To determine the spectrum of bound states and their wave functions
or to determine the scattering parameters and the corresponding
functions for continuous spectrum states, one should solve the
Schr\"{o}dinger equation\vspace*{-1mm}
\begin{equation}
(  \widehat{H}-E)  \Psi_{kL}\left(  r\right)  =0 \label{eq:001}%
\end{equation}
or the Lippmann--Schwinger equation\vspace*{-1mm}
\[\Psi_{kL}\left(  x\right) = \psi_{kL}\left(  x\right)-
\]\vspace*{-8mm}
\begin{equation} \label{eq:002}
- \int G\left( x,\widetilde{x}\right)  \widehat{V}\left(\!
\widetilde{x},\widetilde {\widetilde{x}}\right)  \Psi_{kL}\left(
\widetilde{\widetilde{x}}\right) \
\widetilde{x}^{2}d\widetilde{x}\widetilde{\widetilde{x}}^{2}d\widetilde
{\widetilde{x}}.
\end{equation}
The latter can be written in the coordinate space ($x=r$) or in the
momentum space ($x=p$).\,\,In the coordinate space, we
have\vspace*{-1mm}
\begin{equation}\label{eq:003a}
\psi_{kL}\left(  r\right)=\sqrt{\frac{2}{\pi}}kj_{L}\left(
kr\right)\!,
\end{equation}\vspace*{-7mm}
\begin{equation}\label{eq:003b}
G\left(  r,\widetilde{r}\right)=j_{L}\left(  kr_{<}\right)
n_{L}\left(kr_{>}\right)\!,
\end{equation}\vspace*{-7mm}
\begin{equation}\label{eq:003c}
\widehat{V}\left(  \widetilde{x},\widetilde{\widetilde{x}}\right)
=\delta\left(  \widetilde{r}-\widetilde{\widetilde{r}}\right)
\widehat{V}\left(  \widetilde{r}\right)
\end{equation}
($r_{<}=\min\left(  r,\widetilde{r}\right), \ r_{>}=\max\left(
r,\widetilde{r}\right)  $) and in the momentum space
\begin{equation}\label{eq:003am}
\psi_{kL}\left(  p\right)=\delta\left(  p-k\right),
\end{equation}
\begin{equation}\label{eq:003bm}
G\left(  p,\widetilde{p}\right)=\left(\!  \frac{1}{2}p^{2}-\frac{1}
{2}k^{2}+i\varepsilon\!\right)
^{\!\!-1}\!\delta\left(p-\widetilde{p}\right)\!,
\end{equation}\vspace*{-7mm}
\begin{equation}\label{eq:003cm}
\widehat{V}\left( p,\widetilde{p}\right)=\frac{2}{\pi}p\widetilde
{p}\int\limits_{0}^{\infty}j_{L}\left(  pr\right) \widehat{V}\left(
r\right)j_{L}\left(\widetilde{p}r\right)  r^{2}dr.
\end{equation}
Note that the transition from the coordinate space to the momentum
one is determined by the Fourier--Bessel integral\vspace*{-1mm}
\begin{equation}\label{eq:005}
\Psi_{kL}\left(  p\right)
=\sqrt{\frac{2}{\pi}}p\int\limits_{0}^{\infty}j_{L}\left( pr\right)
\Psi_{kL}\left(  x\right)  r^{2}dr.
\end{equation}

There is another equation, which is also used to determine the
spectrum and the wave functions of bound and scattering
states.\,\,This is the Lippmann--Schwinger equation for the half-off
shell transition T-matrix (see, e.g., \cite{kn:Newton,
kn:Friedrich2013})
\begin{equation}\label{eq:004}
t_{L}\left(  p,k\right)  =V_{L}\left(  p,k\right)
+\int\limits_{0}^{\infty
}d\widetilde{p}\widetilde{p}^{\,2}\frac{V_{L}\left(
p,\widetilde{p}\right)  t_{L}\left(  \widetilde{p},k\right)
}{E-\frac{1}{2}\widetilde{p} ^{2}+i\epsilon}.
\end{equation}
There are several  equivalent definitions of the T-mat\-rix, in
particular, those, which involve integrals with the potential energy
operator and the wave function
\begin{equation}\label{eq:101}
t_{L}\left(  p,k\right)
=\sqrt{\frac{2}{\pi}}p\int\limits_{0}^{\infty}drr^{2} j_{L}\left(
pr\right)  \widehat{V}\left(  r\right)  \Psi_{kL}\left( r\right)
\end{equation}
in the coordinate space or, in the momentum space,
\begin{equation}\label{eq:101a}
t_{L}\left(  p,k\right)
=\int\limits_{0}^{\infty}d\widetilde{p}\widetilde{p}
^{2}\widehat{V}\left( p,\widetilde{p}\right)  \Psi_{kL}\left(
\widetilde {p}\right)\!.
\end{equation}
We present the following important relation, which connects the
T-matrix and the wave function in the momentum space:\vspace*{-3mm}
\begin{equation}
\Psi_{kL}\left(  p\right)  =\delta\left(  p-k\right)  -\frac{t_{L}\left(
p,k\right)  }{\frac{1}{2}p^{2}-\frac{1}{2}k^{2}}. \label{eq:006}%
\end{equation}
By calculating the T-matrix, we can easily construct the wave function of a
system in the momentum space.

It is well known that, in order to determine the spectrum and wave
functions by solving the Schr\"{o}\-din\-ger equation, one needs to
impose the adequate boundary conditions, while the necessary
boundary conditions are included in the Lippmann--Schwinger
equation.

In the present paper, we use the standing-wave representation, which means that the
asymptotic part of wave functions is
\[
\Psi_{kL}^{\left(  a\right)  }\left(  r\right) =\sqrt{\frac{2}{\pi}}
k\ j_{L}\left(  kr\right)  -\tan\delta_{L}\sqrt{\frac{2}{\pi}} k\
n_{L}\left(  kr\right)\!,
\]
where $\delta_{L}$ is the phase shift, and $\ j_{L}\left(  kr\right)
$ and $\ n_{L}\left(  kr\right)  $ are Bessel and Neumann functions,
res\-pec\-ti\-ve\-ly.\,\,Ho\-we\-ver, the T-matrix is usually
determined with the wave function in the running-wave
representation. In this representation,
\[
\Psi_{kL}^{\left(  a\right)  }\left(  r\right)  =\ \left[
\psi_{L}^{\left( -\right)  }\left(  kr\right)  -S_{L}
\psi_{L}^{\left(  +\right) }\left(  kr\right)  \right]\!\!.
\]
Here, $S_{L}=\exp\left\{  2i\delta_{L}\right\}  $ is the scattering
matrix, and
\[
\psi_{L}^{\left(  \pm\right)  }\left(  kr\right)  =\frac{1}{2}\sqrt{\frac
{2}{\pi}}k\left[  -n_{L}\left(  kr\right)  \pm ij_{L}\left(  kr\right)
\right]
\]
are incoming ($\psi_{L}^{\left(  -\right)  }\left(  kr\right)  $) and outgoing
($\psi_{L}^{\left(  +\right)  }\left(  kr\right)  $) waves.

Note that the factor $\sqrt{\frac{2}{\pi}}k$ \ in the definition of
the asymptotic part of a wave function for the continuous spectrum state is
chosen to normalize the wave function $\Psi_{kL}$ by the condition
\begin{equation}\label{eq:103}
\left\langle \Psi_{kL}|\Psi_{\widetilde{k}L}\right\rangle
=\delta\left(\! k-\widetilde{k}\right)\!\!.
\end{equation}

It is easy to show that the T-matrix $T^{\left(  RW\right)  }$
constructed in the running-wave representation is connected with the
T-matrix $T^{\left( SW\right)  }$ in the standing-wave one by
the simple relation
\[
T^{\left(  RW\right)  }=-ie^{i\delta_{L}}\sin\delta_{L} T^{\left(
SW\right)  }.
\]
Thus, we prefer to work with the real function $T^{\left(SW\right)}$.

\subsection{Discrete representation}

To transform the Schr\"{o}dinger equation and the Lippmann--Schwinger equation
for a wave function or T-matrix to the discrete representation, we use a full
set of oscillator functions $\left\{  \Phi_{nL}\left(
x,b\right)\right\}$ in the coordinate space and $\left\{  \Phi_{nL}\left(
p,b\right)  \right\}$ in the momentum space. The explicit form of the functions is as follows:
\begin{equation}\label{eq:107}
\begin{array}{l}
 \Phi_{nL}\left(  r,b\right)=\left(  -1\right)  ^{n}\!\mathcal{N}_{nL} b^{-3/2}\rho^{L}
e^{-\frac{1}{2}\rho^{2}}L_{n}^{L+1/2}\!\left(  \rho^{2}\right)\!, \\[3mm]
\rho=r/b,
\end{array}\!\!\!\!\!\!\!\!
\end{equation}\vspace*{-7mm}
\begin{equation}\label{eq:108}
\begin{array}{l}
\Phi_{nL}\left(  p,b\right)=\mathcal{N} _{nL}
b^{3/2}\rho^{L}e^{-\frac{1}{2}\rho^{2}}L_{n}^{L+1/2}\left(
\rho ^{2}\right)\!,\\[3mm]
\rho =p b.
\end{array}
\end{equation}
Here,\vspace*{-3mm}
\[
\mathcal{N}_{nL}=\sqrt{\frac{2\Gamma\left(  n+1\right)
}{\Gamma\left( n+L+3/2\right)  }}.
\]
These functions obey the completeness relations
\begin{equation}\label{eq:108a}
\sum_{n=0}^{\infty}\Phi_{nL}\left(  r,b\right)  \Phi_{nL}\left(
\widetilde {r},b\right)=\delta\left(  r-\widetilde{r}\right)\!,
\end{equation}\vspace*{-6mm}
\begin{equation}\label{eq:108b}
\sum_{n=0}^{\infty}\Phi_{nL}\left(  p,b\right)  \Phi_{nL}\left(
\widetilde {p},b\right)=\delta\left(  p-\widetilde{p}\right)\!.
\end{equation}

To transform any equation for a wave function or T-matrix to the
oscillator representation, we will use the orthogonality of the
basis functions and the completeness relations.\,\,We can also use
the fact that all quantities, which appear in
Eqs.\,\,(\ref{eq:001}), (\ref{eq:002}), and (\ref{eq:004}), can be
represented as
\begin{subequations}
\label{eq:1088}
\begin{equation}
F\left(  x\right)=\sum_{n=0}^{\infty}\Phi_{nL}\left( x,b\right)
F_{nL}\left(  b\right)\!,
\end{equation}\vspace*{-6mm}
\begin{equation}
\widehat{O}\left(  x\right)=\sum_{n=0}^{\infty}\Phi_{nL}\left(
x,b\right)  \left\langle\! n\left\vert \widehat{O}\right\vert
m\!\right\rangle \Phi_{mL}\left(  x,b\right)\!,
\end{equation}\vspace*{-6mm}
\begin{equation}
\widehat{O}\left(  x,\widetilde{x}\right) =\sum_{n=0}^{\infty}\Phi
_{nL}\left(  x,b\right)  \left\langle\! n\left\vert
\widehat{O}\right\vert m\!\right\rangle \Phi_{mL}\left(
\widetilde{x},b\right)\!,
\end{equation}
\end{subequations}
where $x$ stands for $r$ or $p$.\,\,In Eqs.\,\,(\ref{eq:1088}), it
is tacitly assumed that the function $F\left(  x\right)$, local
operator $\widehat{O}\left(  x\right)$, and nonlocal one
$\widehat{O}\left( x,\widetilde{x}\right)$ obey all necessary
conditions to be expandable in the oscillator basis.

We start the transformation with the Schr\"{o}dinger equation.\,\,It
is easy to verify
that the Schr\"{o}dinger equation in the oscillator representation is%
\begin{equation}\label{eq:112}
\sum_{m=0}^{\infty}\left[  \left\langle\! n\left\vert
\widehat{H}\right\vert m\!\right\rangle -E\delta_{n,m}\right]
C_{m}=0,
\end{equation}
where $\left\{C_{m}\right\}$ are the wave function expansion
coefficients\vspace*{-2mm}
\begin{equation}\label{eq:112a}
\begin{array}{l}
\displaystyle\Psi_{kL}\left( r\right)
=\sum_{m=0}^{\infty}C_{m}\Phi_{nL}\left(
r,b\right)\!, \\[5mm]
\displaystyle C_{m} =\left\langle
\Phi_{nL}|\Psi_{kL}\right\rangle\!.
\end{array}
\end{equation}
Now, we turn our attention to the Lippmann--Schwin\-ger equation for
the wave function.\,\,The integral equation (\ref{eq:002}) is
transformed to a system of linear equations\vspace*{-1mm}
\begin{equation}
C_{n}=C_{n}^{\left(  {\rm B}\right)
}-\sum_{m=0}^{\infty}\sum_{m=0}^{\infty }\left\langle n\left\vert
G\right\vert m\!\right\rangle \left\langle m\left\vert
\widehat{V}\right\vert \widetilde{m}\!\right\rangle C_{n}. \label{eq:113}%
\end{equation}\vspace*{-3mm}

\noindent Here, $\{  C_{n}^{\left(  {\rm B}\right)  }\}  $ is the
wave function of free motion (Bessel function) (\ref{eq:003a}) \ in
the oscillator representation.\,\,Due to the peculiarities of
oscillator functions and the oscillator Hamiltonian, the expansion
coefficients $C_{n}^{\left(  {\rm B}\right)  }$ coincide with
oscillator functions in the momentum space (see more details in
\cite{kn:Heller1,kn:Yamani,kn:Fil_Okhr,kn:SmirnovE})\vspace*{-2mm}
\begin{equation}\label{eq:114}
C_{n}^{\left(  {\rm B}\right)  }=\left\langle \Phi_{nL}|\psi_{kL}\right\rangle
=\Phi_{nL}\left(  p=k,b\right).
\end{equation}\vspace*{-5mm}

\noindent To solve system (\ref{eq:113}), one needs to calculate the
matrix elements of the potential energy operator\vspace*{-1mm}
\[\left\langle\! n\left\vert \widehat{V}\right\vert m\!\right\rangle = \int\limits
_{0}^{\infty}drr^{2}\Phi_{nL}\left(  r,b\right)  \widehat{V}\left(
r\right) \Phi_{mL}\left(  r,b\right)= \]\vspace*{-8mm}
\[= \int\limits_{0}^{\infty}dpp^{2}\Phi_{nL}\left(  p,b\right)  V\left(  p,k\right)
\Phi_{mL}\left(  k,b\right)  k^{2}dk
\]\vspace*{-3mm}

\noindent
 and the matrix elements of
Green's function for the free motion Hamiltonian between oscillator
functions\vspace*{-1mm}
\[\left\langle n\left\vert G\right\vert m\right\rangle =\]\vspace*{-9mm}
\[= \int\limits_{0}^{\infty
}dpp^{2}\Phi_{nL}\left(  p,b\right)  \left[
E-\frac{1}{2}p^{2}+i\epsilon \right]  ^{-1}\Phi_{mL}\left(
p,b\right)=\]\vspace*{-7mm}
\begin{equation}\label{eq:119}
= \int\limits_{0}^{\infty}\int\limits_{0}^{\infty}drr^{2}\Phi_{nL}\left(  r,b\right)
G\left(  r,\widetilde{r}\right)  \Phi_{mL}\left(  \widetilde{r},b\right)
\widetilde{r}^{2}d\widetilde{r}.
\end{equation}\vspace*{-3mm}

\noindent In Ref.\,\,\cite{Heller1975}, one can find the explicit
form of the matrix elements $\left\langle n\left\vert G\right\vert
m\right\rangle $ and the recurrence relations they satisfy.

There are two different ways to present the T-mat\-rix in the
discrete (oscillator) form.\,\,First, we can use the
expansion\vspace*{-2mm}
\begin{equation}\label{eq:106}
t_{L}\left(  p,k\right)  =\sum_{n=0}^{\infty}\Phi_{nL}\left(  p,b\right)
t_{nL}\left(  b,k\right)\!.
\end{equation}\vspace*{-3mm}

\noindent It is obvious that the expansion coefficients
$\left\{t_{nL}\left(b,k\right) \right\}$ are determined
as\vspace*{-2mm}
\begin{equation}\label{eq:106a}
t_{nL}\left(  b,k\right)  =\int\limits_{0}^{\infty}dpp^{2}\Phi_{nL}\left(
p,b\right)  t_{L}\left(  p,k\right)\!.
\end{equation}\vspace*{-5mm}

\noindent Thus, we have to deal with an infinite vector.

Second, we can represent the T-matrix as a matrix\vspace*{-1mm}
\begin{equation}\label{eq:109}
t_{L}\left(  p,k\right)  =\sum_{n,m=0}^{\infty}\Phi_{nL}\left(  p,b\right)
t_{nm}\left(  b\right)  \Phi_{mL}\left(  k,b\right)\!,
\end{equation}\vspace*{-3mm}

\noindent where the matrix elements $t_{nm}\left(  b\right)  $ are
determined as\vspace*{-2mm}
\[
t_{nm}\left(  b\right)
=\int\limits_{0}^{\infty}dpp^{2}\Phi_{nL}\left(  p,b\right)
t_{L}\left(  p,k\right) \times
\]\vspace*{-7mm}
\begin{equation}\label{eq:109a}
\times\, \Phi_{mL}\left(  k,b\right)  k^{2}dk.
\end{equation}\vspace*{-5mm}

\noindent We note that the expansion coefficients for the T-mat\-rix
in both representations depend on the oscillator length.\,\,In the
next section, we will study how strongly the T-matrix expansion
coefficients depend on the oscillator length $b$.

By projecting Eq.\,\,(\ref{eq:004}) on the oscillator basis, we
obtain the sets of linear inhomogeneous equations for the vector
$t_{n}=t_{nL}(b,k)$,\vspace*{-2mm}
\begin{equation}\label{eq:110}
t_{n}=V_{n}\left(  b,k\right)
+\sum_{m,\widetilde{m}=0}^{\infty}\left\langle\! n\left\vert
\widehat{V}\right\vert m\!\right\rangle \left\langle m\left\vert
G\right\vert \widetilde{m}\right\rangle t_{\widetilde{m}}
\end{equation}\vspace*{-7mm}

\noindent or\vspace*{-2mm}
\begin{equation}\label{eq:110a}
\sum_{\widetilde{m}=0}^{\infty}\left[
\delta_{n\widetilde{m}}\!-\!\sum _{m=0}^{\infty}\left\langle\!
n\left\vert \widehat{V}\right\vert m\!\right\rangle \left\langle
m\left\vert G\right\vert \widetilde{m}\right\rangle \right]
t_{\widetilde{m}}\!=\!V_{n}\left(  b,k\right),
\end{equation}\vspace*{-7mm}

\noindent and for the matrix $t_{nm}=t_{nm}\left(
b\right)$:\vspace*{-1mm}
\begin{equation}\label{eq:111}
t_{nm}=V_{nm}+\sum_{m,\widetilde{m}=0}^{\infty}\left\langle\!
n\left\vert \widehat{V}\right\vert m\!\right\rangle \left\langle
m\left\vert G\right\vert \widetilde{m}\right\rangle
t_{\widetilde{m}m}
\end{equation}\vspace*{-5mm}

\noindent or\vspace*{-3mm}
\begin{equation}\label{eq:111a}
\sum_{\widetilde{m}=0}^{\infty}\left[  \delta_{n\widetilde{m}}-\sum
_{m=0}^{\infty}\left\langle\! n\left\vert \widehat{V}\right\vert
m\right\rangle \!\left\langle m\left\vert G\right\vert
\widetilde{m}\right\rangle \right] t_{\widetilde{m}m}=V_{nm}.
\end{equation}\vspace*{-7mm}

\noindent Here,\vspace*{-3mm}
\[V_{nL}\left(  b,k\right)  =\sqrt{\frac{2}{\pi}}k\int\limits_{0}^{\infty}
drr^{2}\Phi_{nL}\left(  r,b\right)  \widehat{V}\left(  r\right)
j_{L}\left( kr\right)=\]\vspace*{-7mm}
\[=\int\limits_{0}^{\infty}dpp^{2}\Phi_{nL}\left(  p,b\right)  V_{L}\left(
p,k\right)  =\sum_{m=0}^{\infty}\left\langle\! n\left\vert \widehat
{V}\right\vert m\!\right\rangle C_{m}^{\left(  B\right)  }.\]

Let us consider the T-matrix expansion coefficients
(\ref{eq:106a}).\,\,By using the definition of the T-matrix
(\ref{eq:101}), this equation can be rewritten as\vspace*{-1mm}
\[t_{nL}\left(  b,k\right)  = \left\langle\! \Phi_{nL}\left\vert \widehat
{V}\right\vert \Psi_{kL}\!\right\rangle =\]
\begin{equation}\label{eq:120}
=\int\limits_{0}^{\infty}drr^{2}\Phi
_{nL}\left(  r,b\right)  \widehat{V}\Psi_{kL}\left(  r\right)
\end{equation}
or\vspace*{-3mm}
\begin{equation}\label{eq:121}
t_{nL}\left(  b,k\right)  =\sum_{m=0}^{\infty}\left\langle\!
n\left\vert \widehat{V}\right\vert m\!\right\rangle C_{m}.
\end{equation}
Here, $\left\{  C_{m}\right\}  $ are expansion coefficients of the
wave function $\Psi_{kL}\left(  r\right)$ (see
Eq.\,\,(\ref{eq:112a})).

\subsection{Square-well potential}

To check our numerical results, we need to consider a potential,
which admits a simple expression for the T-matrix.\,\,There is only
a restricted number of two-body problems, which can be solved
analytically, and for which the T-matrix can be obtained in a closed
analytic form.\,\,For instance, the T-matrix in Ref.
\cite{1982PhRvC..26.2381K} was constructed for a delta-shell
potential.

One of these cases is the square-well potential\vspace*{-1mm}
\[
\widehat{V}\left(  r\right)  =\left\{\!\!
\begin{array}
{ll}%
V_{0} & r\leq a,\\[1mm]
0 & r>a.
\end{array}
\right.
\]
This case has been considered in detail many times. See, for
instance, Ref.\,\,\cite{1984JPhG...10.1639D}, where the off-shell
T-mat\-rix was obtained.\,\,We also consider this potential in
detail, assuming that $V_{0}$ is negative, and the potential is
attractive. The obtained results will be intensively used in Section
\ref{sec:Analysis}.

To determine the T-matrix for a square-well potential, we have to
obtain the wave function.\,\,The wave function for the potential in
the internal region ($0\leq r\leq a$) reads\vspace*{-3mm}
\begin{equation}\label{eq:201}
\Psi_{kL}\left(  r\right)  =\Psi_{kL}^{\left(  i\right)  }\left(  r\right)
=\mathcal{A}_{kL}\sqrt{\frac{2}{\pi}}kj_{L}\left(  k_{0}r\right)\!,
\end{equation}
where $\mathcal{A}_{k}$ is the normalization constant,
and\vspace*{-1mm}
\begin{equation}\label{eq:201a}
k_{0}=\sqrt{2\left(  E+V_{0}\right)  },\quad k=\sqrt{2E}.
\end{equation}
In the asymptotic region ($r\geq a$) for the continuous-spectrum
states, we have\vspace*{-1mm}
\begin{equation}\label{eq:202}
\Psi_{kL}^{\left(  a\right)  }\left(  r\right)\!=\!\sqrt{\frac{2}{\pi}}
k\ j_{L}\left(  kr\right)\!-\!\tan\delta_{L}\sqrt{\frac{2}{\pi}}
k\ n_{L}\left(  kr\right)\!.
\end{equation}
By matching the internal  (\ref{eq:201}) and asymptotic
(\ref{eq:202}) parts of the wave function and their first
derivatives as well, we can determine the phase shift $\delta_{L}$
and the constant $\mathcal{A}_{kL}$.

The T-matrix for the square-well potential is fully determined by
the internal part\ (\ref{eq:201}) of the wave function\vspace*{-2mm}
\begin{equation}\label{eq:204}
t_{L}\!=\!t_{L}\left(
p,k\right)\!=\!\sqrt{\frac{2}{\pi}}V_{0}\mathcal{A}_{kL}
\!\!\int\limits_{0}^{a}\!\!drr^{2}j_{L}\left(  pr\right)
j_{L}\left( k_{0}r\right)\!.
\end{equation}
It is easy to verify that, for the zero value of orbital momentum
$L=0,$
\[t_{0}\left(  p,k\right) =\frac{2}{\pi}\frac{k}{k_{0}}V_{0}\mathcal{A}_{k}\,\times\]\vspace*{-5mm}
\begin{equation}\label{eq:205}
\times\,\frac{p\cos\left(  pa\right)  \sin\left(  k_{0}a\right)
-k_{0} \sin\left(  pa\right)  \cos\left(  k_{0}a\right)
}{k_{0}^{2}-p^{2}}
\end{equation}
and
\[
A_{k}=k/\left[  k\sin\left(  k_{0}a\right)  \sin\left(  ka\right)
+k_{0} \cos\left(  k_{0}a\right)  \cos\left(  ka\right)  \right]\!.
\]
On the energy shell, we have
\begin{equation}\label{eq:206}
t\left(  k,k\right) =\frac{2}{\pi}\frac{k^{2}}{k_{0}}\tan\delta_{0}.
\end{equation}

Note that expression (\ref{eq:205}) represents the T-matrix not only for
scattering states ($E\geq0$), but also for bound state(s) ($-V_{0}\leq E<0$).
One has to calculate the energy of a bound state (or the momentum $k_{0}$) and
the corresponding normalization factor $A_{k}$.

\section{Analysis of Results}
\label{sec:Analysis}

The numerical analysis of the T-matrix will be carried out for the $s$-state ($L=0$)
only, where the interaction is more stronger than in other orbital states ($L>0$)
 and is not diminished by the centrifugal barrier.\,\,Four potentials are used
to study properties of the T-matrix.\,\,They are Gaussian (G),
exponential (E), Yukawa (Y), and square-well (SW) potentials:
\begin{align}
\widehat{V}(r)&=V_{0}\exp\left\{\!  -\left(  \frac{r}
{a}\right)  ^{\!\!2}\!\right\}\!, \quad\text{(G)}\nonumber\\
\widehat{V}(r)&=V_{0}\exp\left\{\!  -\frac{r}{a}\!\right\}\!
,\quad\text{(E)}\label{eq:301}\\
\widehat{V}(r)&=V_{0}\exp\left\{\!  -\frac{r}{a}\!\right\}
\left(  \frac{a}{r}\right)\!, \quad\text{(Y)}\nonumber\\
\widehat{V}(r)&=\left\{\!\!
\begin{array}
{ll}%
V_{0} & r\leq a,\\[2mm]
0 & r>a,
\end{array}
\right.   \quad\text{(SW)}\nonumber
\end{align}
In this section, we use the nuclear units for energy (MeV) and
length (fm), so that the constant $\hbar^{2}/m=$ $=41.47$
MeV$\cdot$\,fm$^{2}$.

\begin{figure}%
\includegraphics[width=\column]{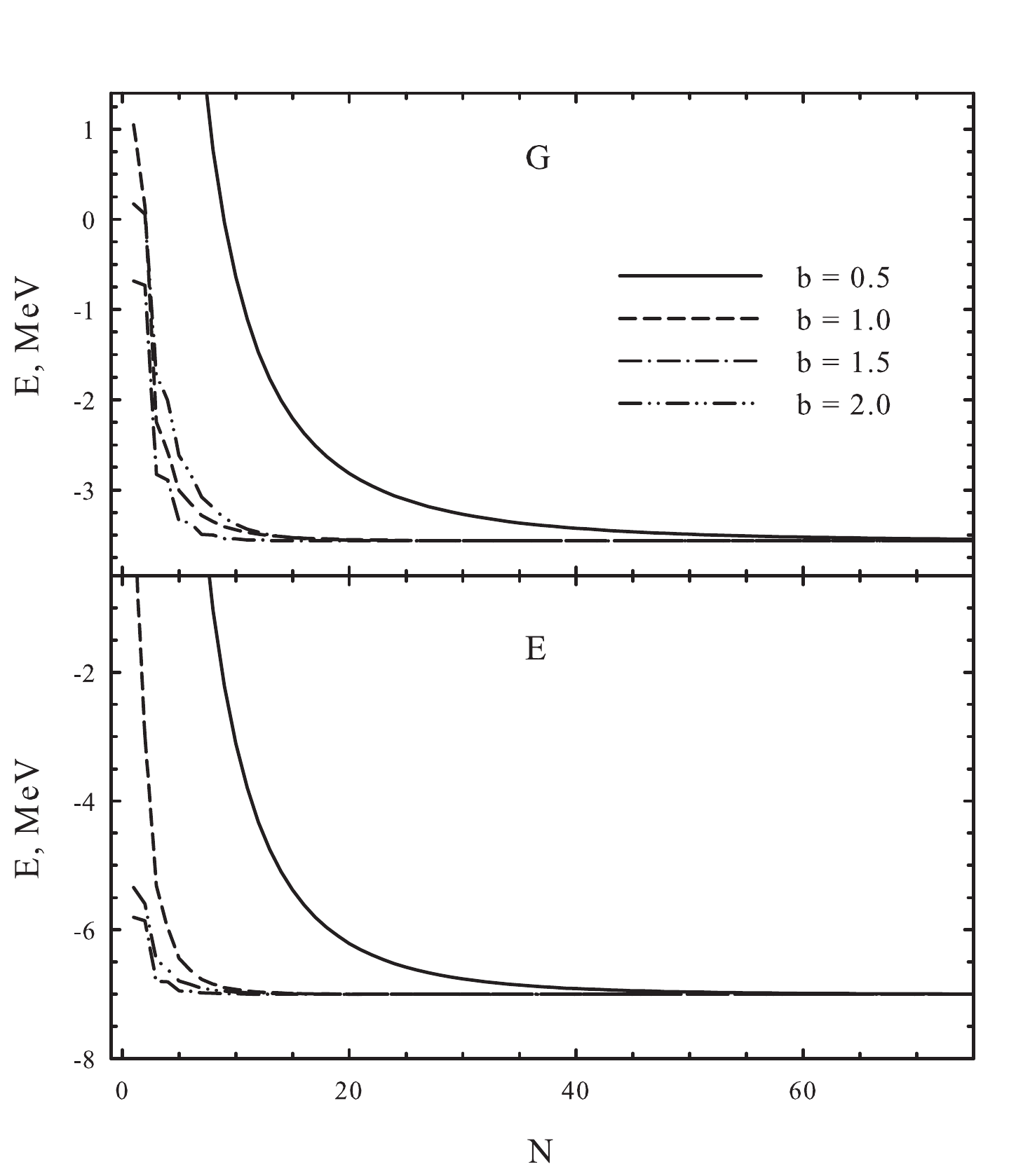}
\vskip-3mm\caption{ Convergence of the ground-state energy for the
Gauss and exponential potentials
}\label{Fig:GRState_G&E}\vspace*{-2mm}
\end{figure}

For all potentials, we take the radius of the potential $a=1$ fm,
and the depth $V_{0}=-85$ MeV.\,\,With such a choice of parameters,
we obtain one bound state for all potentials, whose energies are
listed in Table.\,\,It was established by the numerical solution of
the set of equations (\ref{eq:112}) with a maximal number of
oscillator functions (see the text bellow).\,\,One can see that we
obtained a deeply bound state for the Yukawa potential, a weakly
bound state for the Gaussian potential, and a moderately bound state
for the exponential and square-well potentials.\,\,The results
demonstrated in Table are obtained with $N=300$ functions and the
oscillator length $b=a=1.0$ fm.

\begin{table}[b]
\vspace*{-1mm} \noindent\caption{ Energy of the bound state
}\vskip3mm\tabcolsep8.9pt

\noindent{\footnotesize\begin{tabular}{|c|c|c|c|c|}
  \hline
\multicolumn{1}{|c}{\rule{0pt}{5mm}Potential} &
\multicolumn{1}{|c}{G } & \multicolumn{1}{|c}{E } &
\multicolumn{1}{|c}{Y } &
\multicolumn{1}{|c|}{SW } \\[1.5mm]
  \hline \rule{0pt}{4mm}$E$, MeV & $-3.564$ & $-7.006$ & $-26.744$ & $-9.388$\\[2mm]
  \hline
\end{tabular}}
\end{table}

In what follows, we will use four values of the oscillator radius
$b=0.5$, 1.0, 1.5, and 2.0 fm to study the dependence of the
T-matrix expansion coefficients on the oscillator
radius.\,\,Calculations are organized in the following
way.\,\,First, we construct the $N\times N$ matrix of the
Hamiltonian, where 1$\leq N\leq 500$.\,\,(We do not dwell here on
the calculations of matrix elements of the potential energy operator
between oscillator functions.\,\,We make use of the technique of
generating functions, details of which can be found in Refs.
\cite{kn:cohstate1E, kn:cohstate2E}).\,\,Second, we calculate the
eigenvalues (spectrum) and the eigenfunctions of the matrix by using
the discrete form of the Schr\"{o}dinger equation
(\ref{eq:112}).\,\,Then we obtain the T-matrix of bound and
pseudobound states in the oscillator representation, by using
Eq.\,\,(\ref{eq:121}).\,\,Third, we solve the  system of linear
equation (\ref{eq:113}), which determine the wave function and the
phase shift of the scattering state with fixed energy $E,$ and then
we construct the T-matrix by using
Eq.\,\,(\ref{eq:121}).\,\,Finally, on the fourth step, we solve the
set of linear equations (\ref{eq:110}) and obtain directly the
T-matrix for a fixed energy $E$.\,\,We make use of the third way to
check the correctness of the calculations of the T-matrix in the
fourth way.

In this section for the sake of convenience, we will denote the
T-matrix as $t\left(  p,E\right)$ explicitly indicating the energy
of the discrete or continuous-spectrum state.\vspace*{-1mm}

\subsection{Convergence}

\begin{figure}%
\vskip1mm
\includegraphics[width=\column]{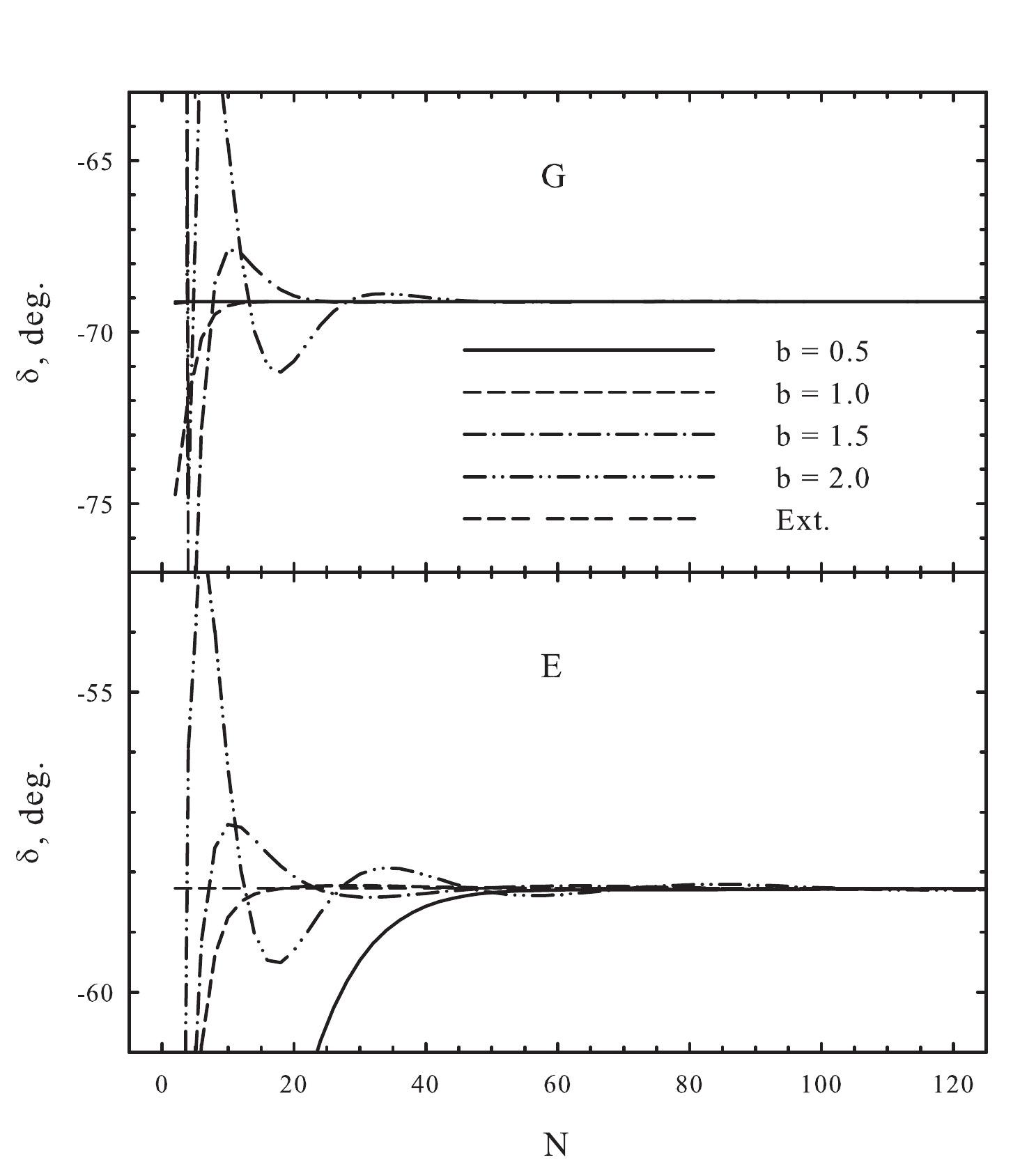}
\vskip-3mm\caption{ Phase shift as a function of the number of
oscillator functions involved in calculations.\,\,Results are
obtained for the Gauss and exponential potentials with the energy
$E=5.0$ }\label{Fig:PhaseConv_G&E}\vspace*{2mm}
\end{figure}

First of all, we consider whether the oscillator basis is large
enough to provide the convergent results for the bound state energy
and the phase shift of continuous-spectrum states.\,\,In
Fig.\,\,\ref{Fig:GRState_G&E}, we show the dependence of the bound
state energy $E$ on the number of oscillator functions $N$ involved
in calculations.\,\,These results are obtained for two potentials
(Gaussian and exponential) and for four different values of
oscillator length $b$.\,\,One can see that one hundred functions
give a very stable value of ground state energy for all oscillator
lengths.\,\,The dependence of the phase shift on the number of
oscillator functions $N$ used in calculations is demonstrated in
Fig.\,\,\ref{Fig:PhaseConv_G&E}.\,\,The phase shift is determined
for the energy $E= 5.0$ MeV.\,\,The exact values of phase shift are
calculated within the variable phase method \cite{kn:Calogero,
kn:Babikov}.\,\,To obtain the stable phase shift independent of $N$,
we need to use more oscillator functions than for bound state
calculations.\vspace*{-1mm}

\subsection{Ground state}

\begin{figure}
\vskip1mm
\includegraphics[width=\column]{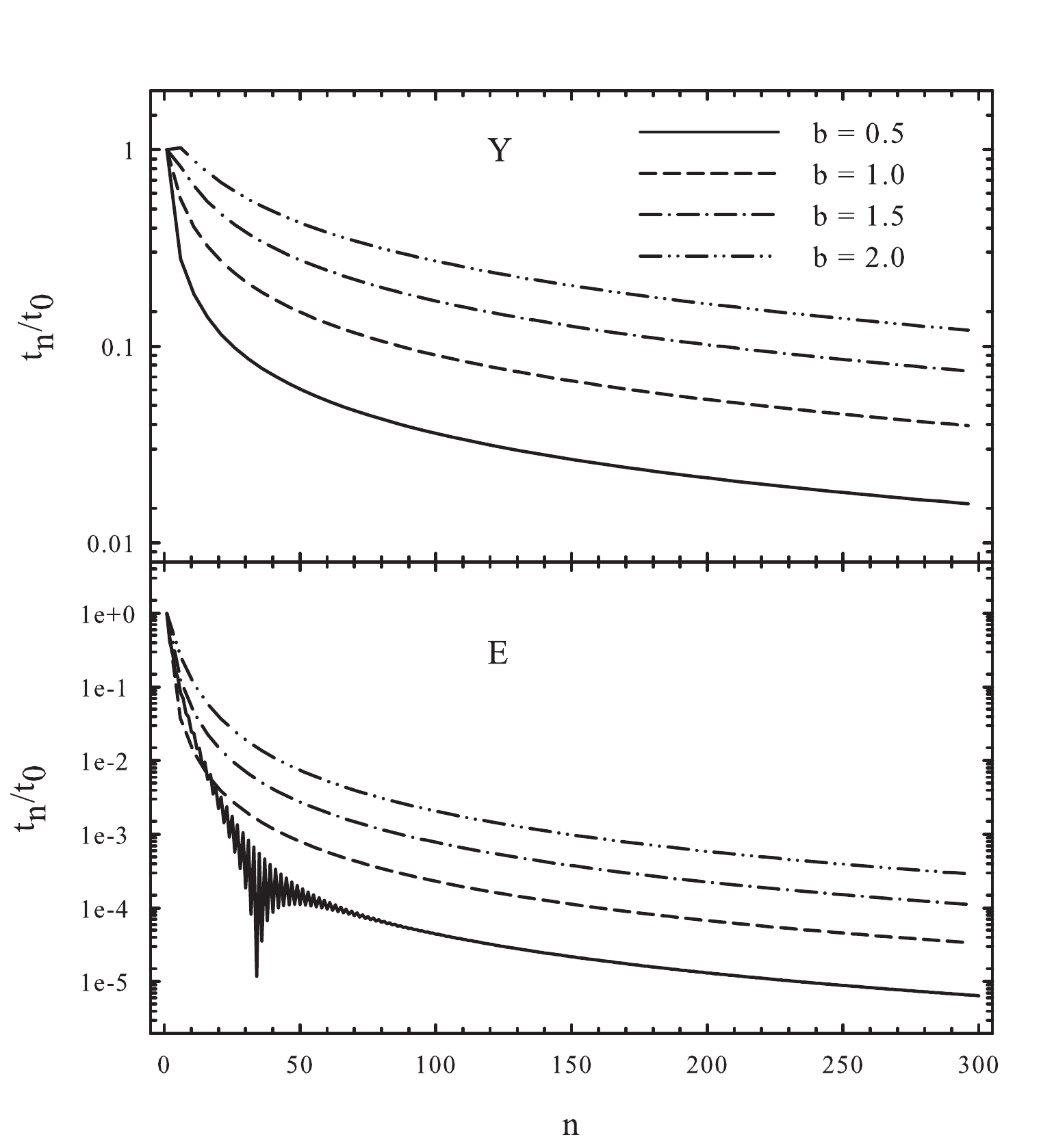}
\vskip-3mm\caption{ T-matrix for bound states calculated with
the Yukawa and exponential potentials }\label{Tmatr_BS_Y&E}
\end{figure}

In this section, we consider the T-matrix of bound states.\,\,In
Fig.\,\,\ref{Tmatr_BS_Y&E}, we show the T-matrix for bound states
with the Yukawa and exponential potentials. To demonstrate the rate
of decreasing of the T-matrix expansion coefficients, we display the
renormalized expansion coefficients $t_{n}/t_{0}=t_{n}\left(
b,E\right)  /t_{0}\left(  b,E\right)  $. For the Yukawa potential,
the bound state is a deeply bound one with its energy to be $-26.74$
MeV.\,\,This explains why the T-matrix for this potential decreases
very rapidly, as $n$ increases.\,\,The common feature of the
T-matrix for the Yukawa and exponential potentials is that the
larger the oscillator length $b$, the slower is the decreasing of
the T-matrix expansion coefficients.\,\,The same tendency is
observed for the Gaussian and square-well potentials.

\begin{figure}%
\vskip1mm
\includegraphics[width=\column]{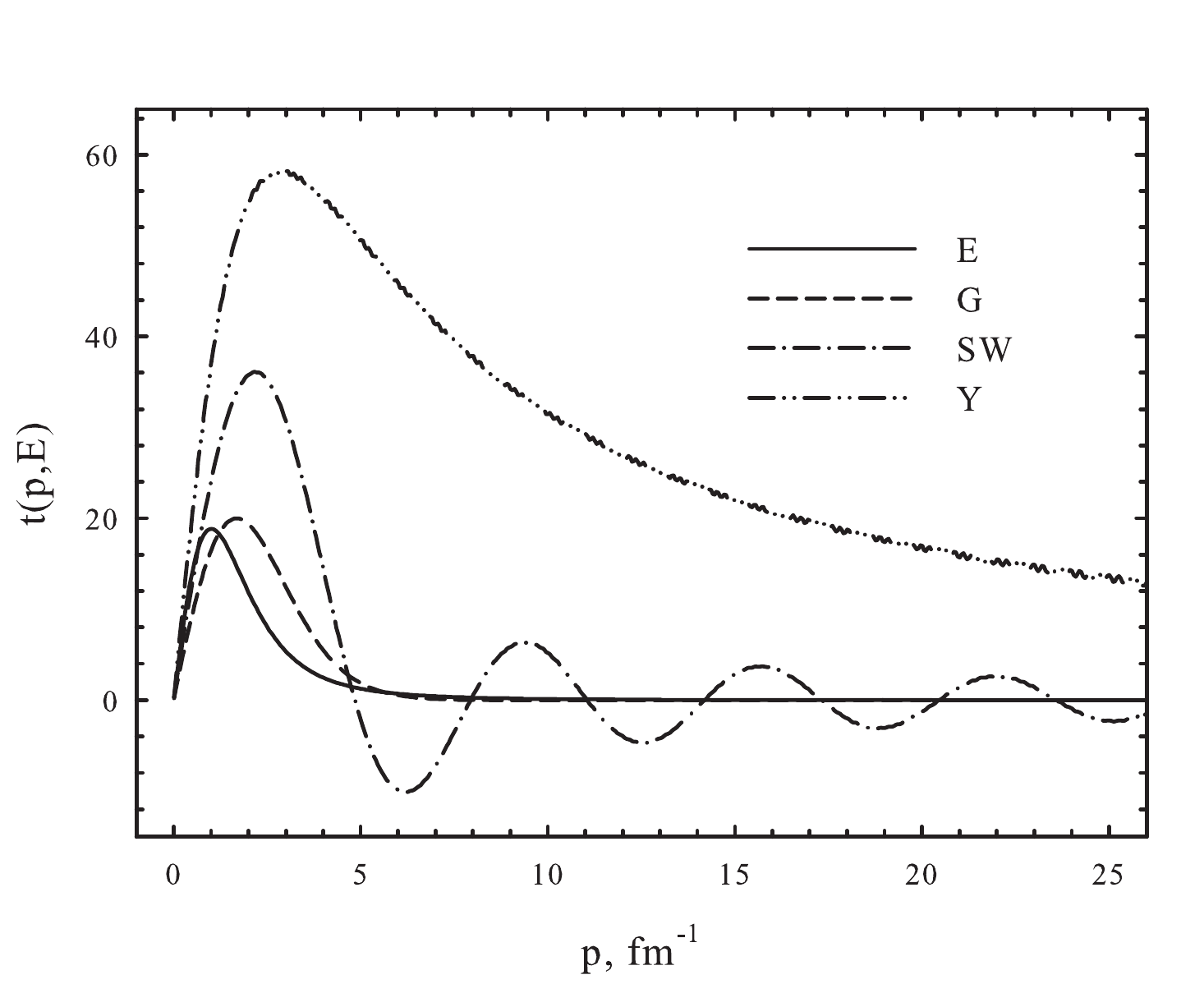}
\vskip-3mm\caption{ T-matrix of the bound state as a function of the
momentum $p$ calculated with four potentials
}\label{Fig;Tmatr_GS_MS}\vspace*{-2mm}
\end{figure}

\begin{figure}%
\vskip1mm
\includegraphics[width=\column]{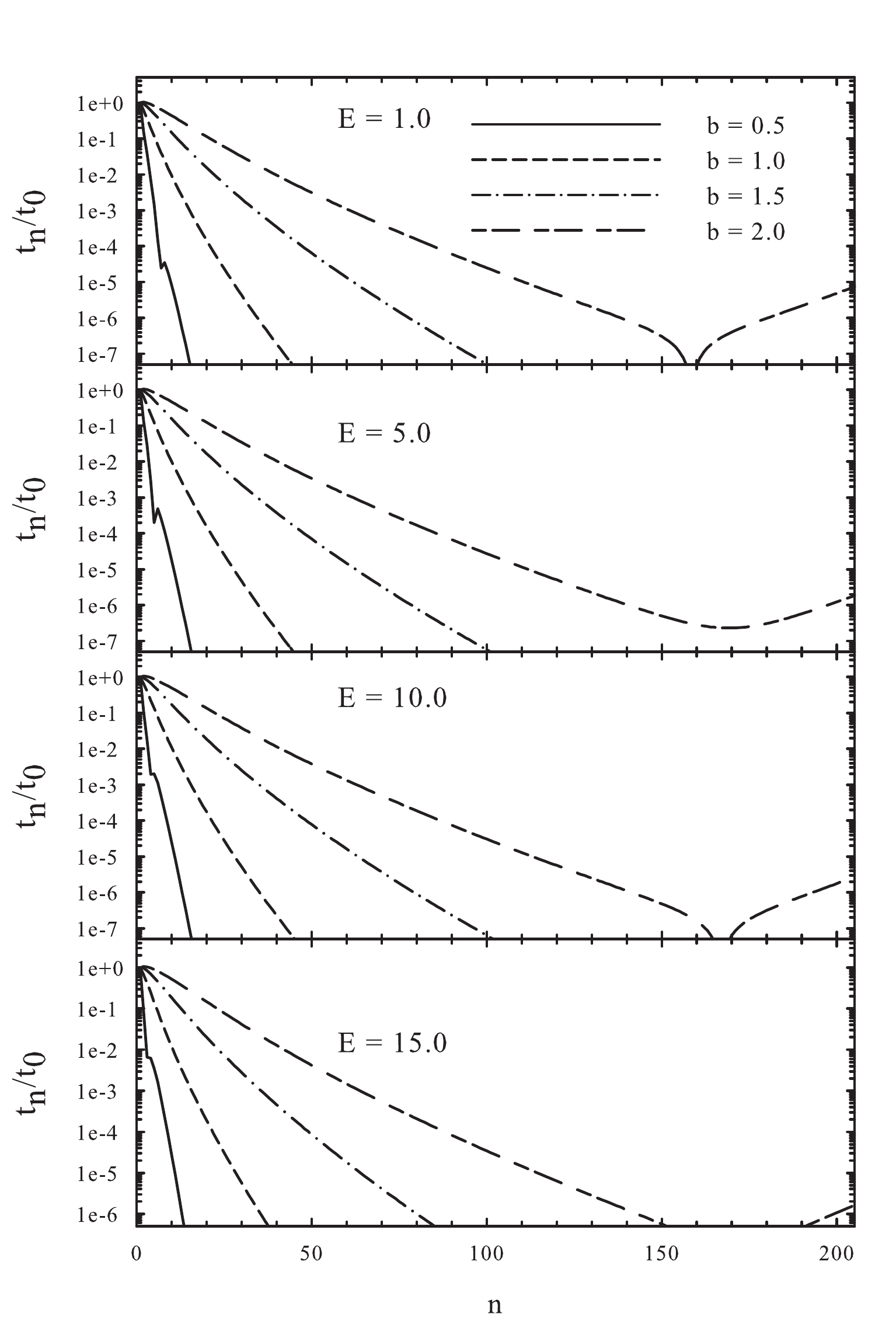}
\vskip-3mm\caption{ T-matrix expansion coefficients for
continuous-spect\-rum states obtained with the Gaussian potential
}\label{Fig:Tmatr_G_EvsbN}\vspace*{-2mm}
\end{figure}

\begin{figure}%
\vskip1mm
\includegraphics[width=\column]{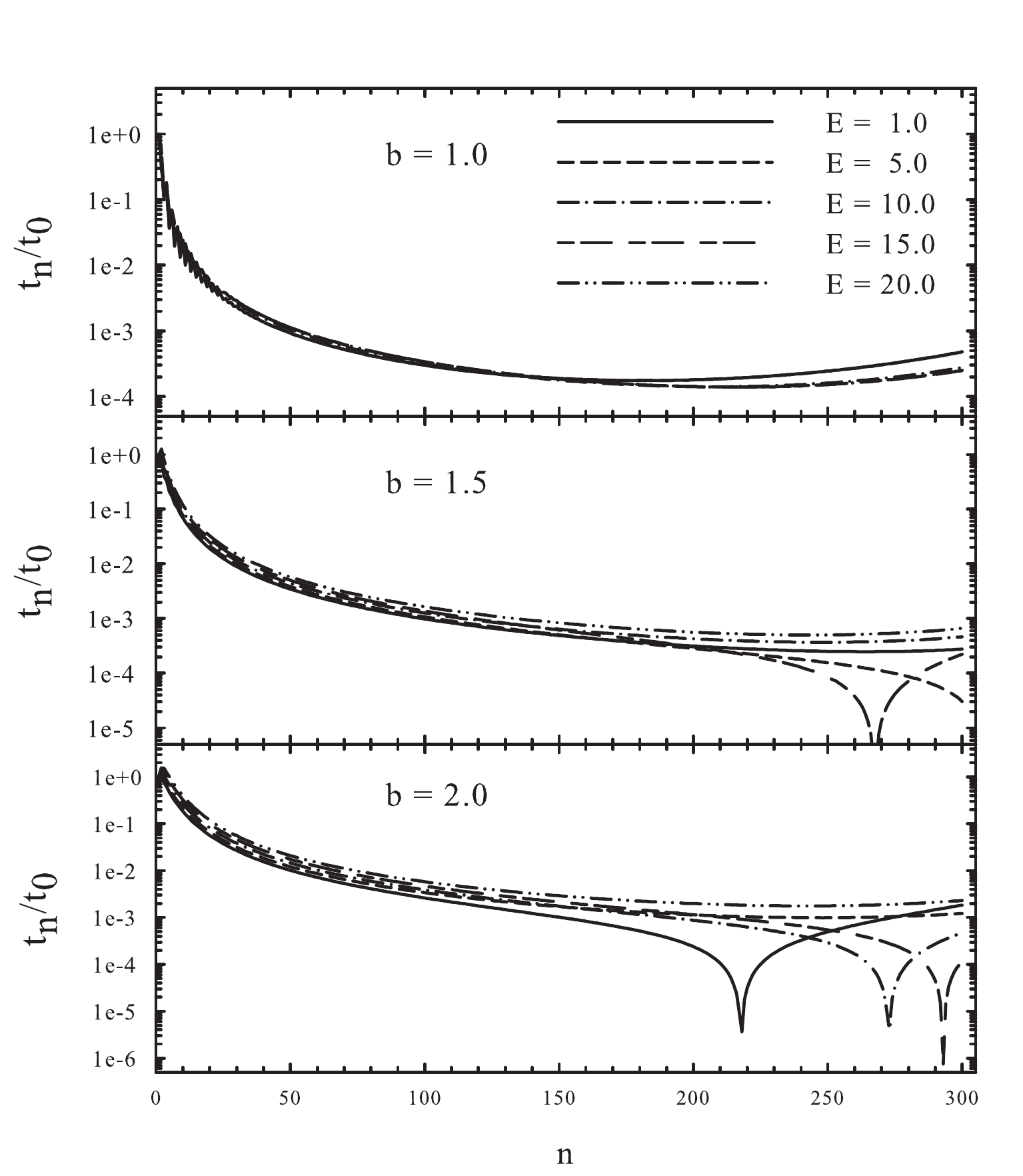}
\vskip-3mm\caption{ The T-matrix expansion coefficients as a
function of $n$ calculated with the exponential potential
}\label{Fig:Tmatr_E_EvsbNS}\vspace*{-2mm}
\end{figure}

\begin{figure}%
\vskip1mm
\includegraphics[width=\column]{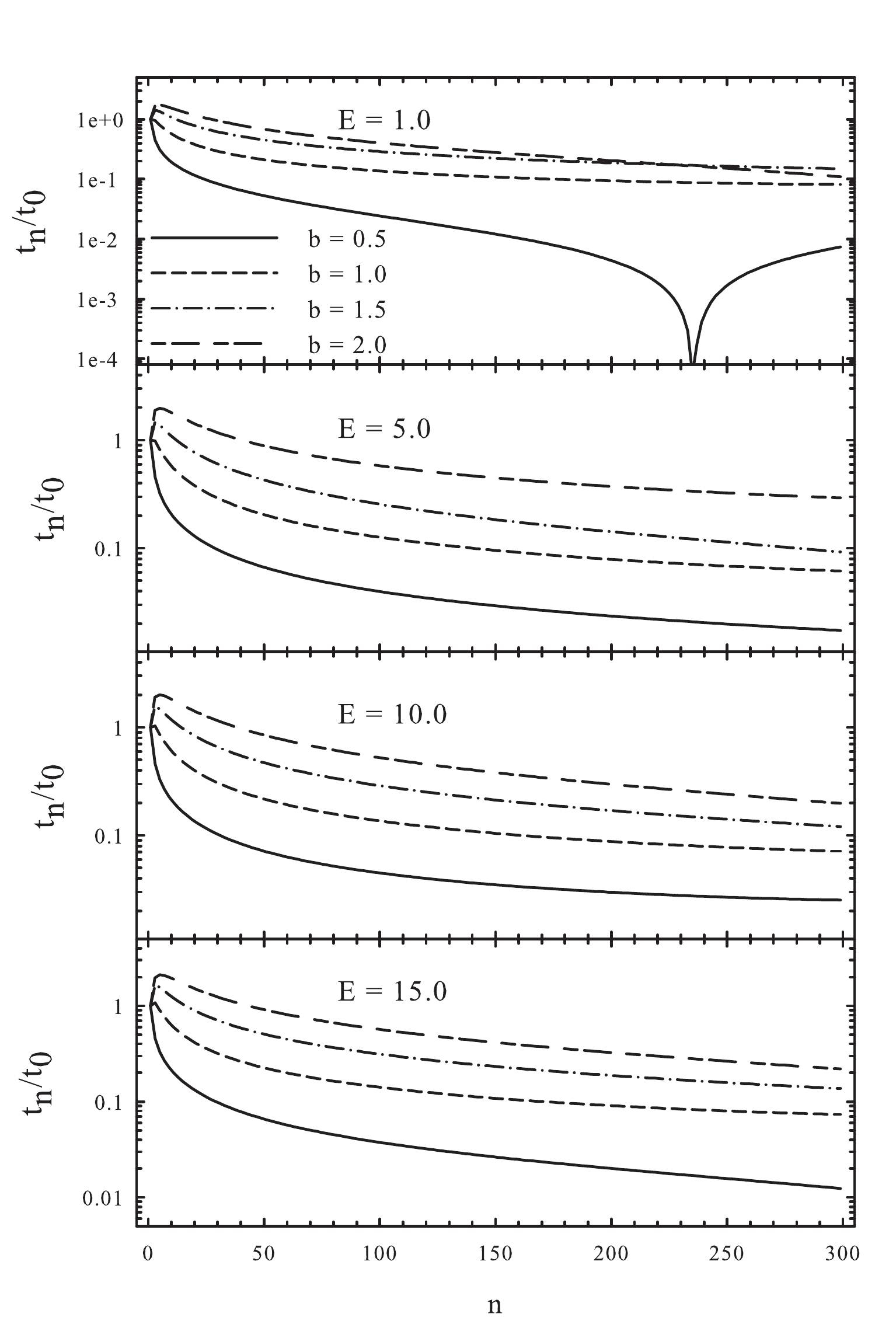}
\vskip-3mm\caption{ Expansion coeficients $t_{n}/t_{0}$, obtained
with the Yukawa potential }\label{Fig:Tmatr_Y_EvsbN}\vspace*{-2mm}
\end{figure}

With the help of relation (\ref{eq:106}), we can easily construct
the T-matrix of a bound state in the momentum space.\,\,In Fig.
\ref{Fig;Tmatr_GS_MS}, we display the T-matrix $t\left(  p,E\right)$
as a function of the momentum $p$ for the bound state for four
potentials.\,\,These results are obtained with the oscillator length
$b=1$ fm and with $N=300$ oscillator functions.\,\,However, by using
other values of oscillator length, we obtain the same results.\,\,It
is worth to note that the T-matrix for a deeply bound state (Yukawa
potential) is very dispersed in the momentum space.\,\,This reflects
the fact that the behavior of the T-matrix in the momentum space is
determined by the wave function in the internal and asymptotic
regions (see Eq.\,\,(\ref{eq:101})).\,\,In some cases, the internal
part of wave functions gives a larger contribution than the
asymptotic part.\,\,The same is true for the square-well potential,
where the T-matrix is fully determined by the internal part of the
wave function (as was pointed out above (see
Eq.\,\,(\ref{eq:204}))).\,\,Contrary to the case of the Yukawa and
square-well potentials, the T-matrix for the Gaussian and
exponential potentials is represented by the small values of
momentum $p$: $0\leq p\leq7$ fm$^{-1}$.\vspace*{-1mm}

\subsection{Continuous-spectrum states}

We selected five values of energy ($E=1$, 5, 10, 15, and 20 MeV) to
study the dependence of the T-matrix expansion coefficients on the
energy in the continuous spectrum.\,\,The energy range $0<E\leq 20$
MeV selected in our calculations represents the region, where the
effect of the potential energy is stronger than in the high-energy
region.

As we are interested in the study of the rate of decreasing of
$t_{n}$, we will display the renormalized expansion coefficients
$\overline{t}_{n}=t_{n}/t_{0}$.\,\,In
Fig.\,\,\ref{Fig:Tmatr_G_EvsbN}, we show the T-matrix expansion
coefficients for the Gauss potential for four values of energy
$E=1$, 5, 10, and 15\,\,MeV. One can see that $\overline{t}_{n}$
drops to zero very rapidly, as $n$ is increased.\,\,The smaller the
oscillator length, the faster is the decreasing of the T-matrix
expansion coefficients for the Gauss potential.\,\,Note that the
rate of decreasing of $\overline{t}_{n}$ almost independent of the
energy used in our calculations.

\begin{figure}%
\vskip1mm
\includegraphics[width=\column]{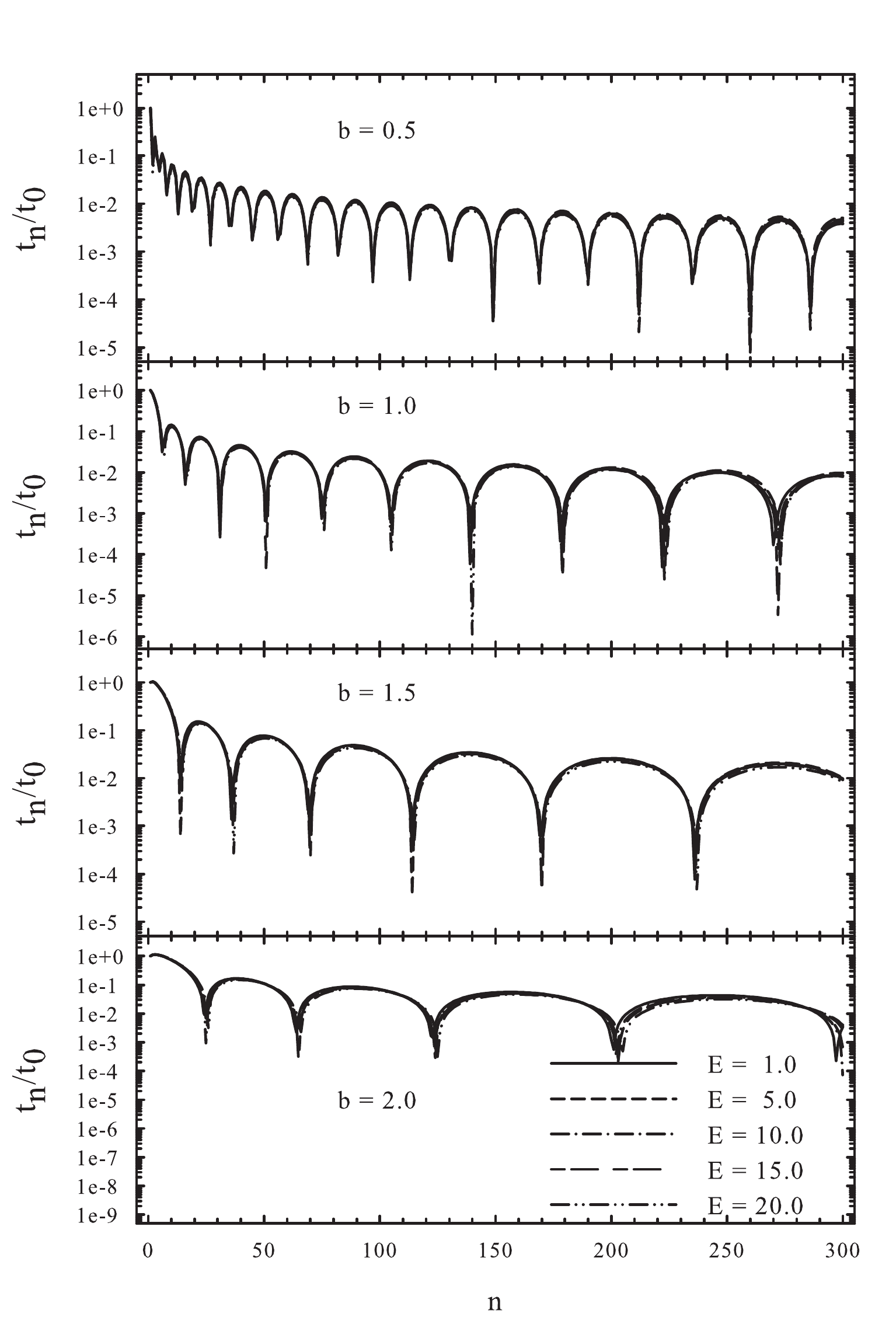}
\vskip-3mm\caption{ Square-well potential.\,\,The renormalized
T-matrix ex\-pan\-sion coefficients as a function of $n$
}\label{Fig:Tmatr_SW_EvsbN}\vspace*{-3mm}
\end{figure}

By comparing Fig.\,\,\ref{Fig:Tmatr_G_EvsbN} with Figs.
\ref{Fig:Tmatr_E_EvsbNS}, \ref{Fig:Tmatr_Y_EvsbN}, and
\ref{Fig:Tmatr_SW_EvsbN}, we can see that the fast decrease of
expansion coefficients is observed only for the Gaussian potential,
while the other potentials exhibit a much more slower decrease.

\begin{figure}%
\vskip1mm
\includegraphics[width=\column]{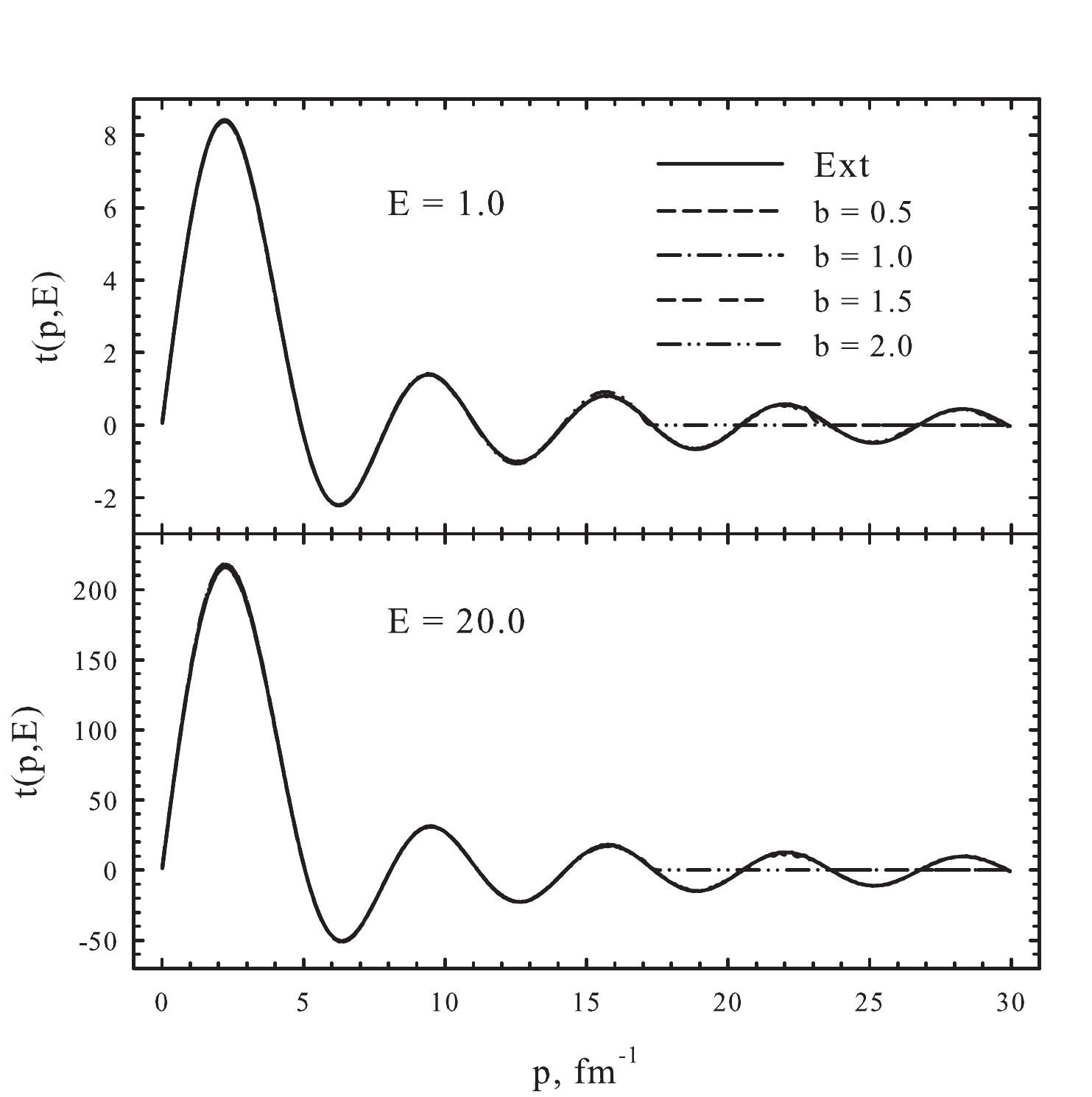}
\vskip-3mm\caption{ T-matrix convergence for the square-well potential
calculated for two values of energy $E=1$ and $E=10$ MeV
}\label{Fig:Tmatrix_SW_MS}
\end{figure}

\begin{figure}%
\vskip3mm
\includegraphics[width=\column]{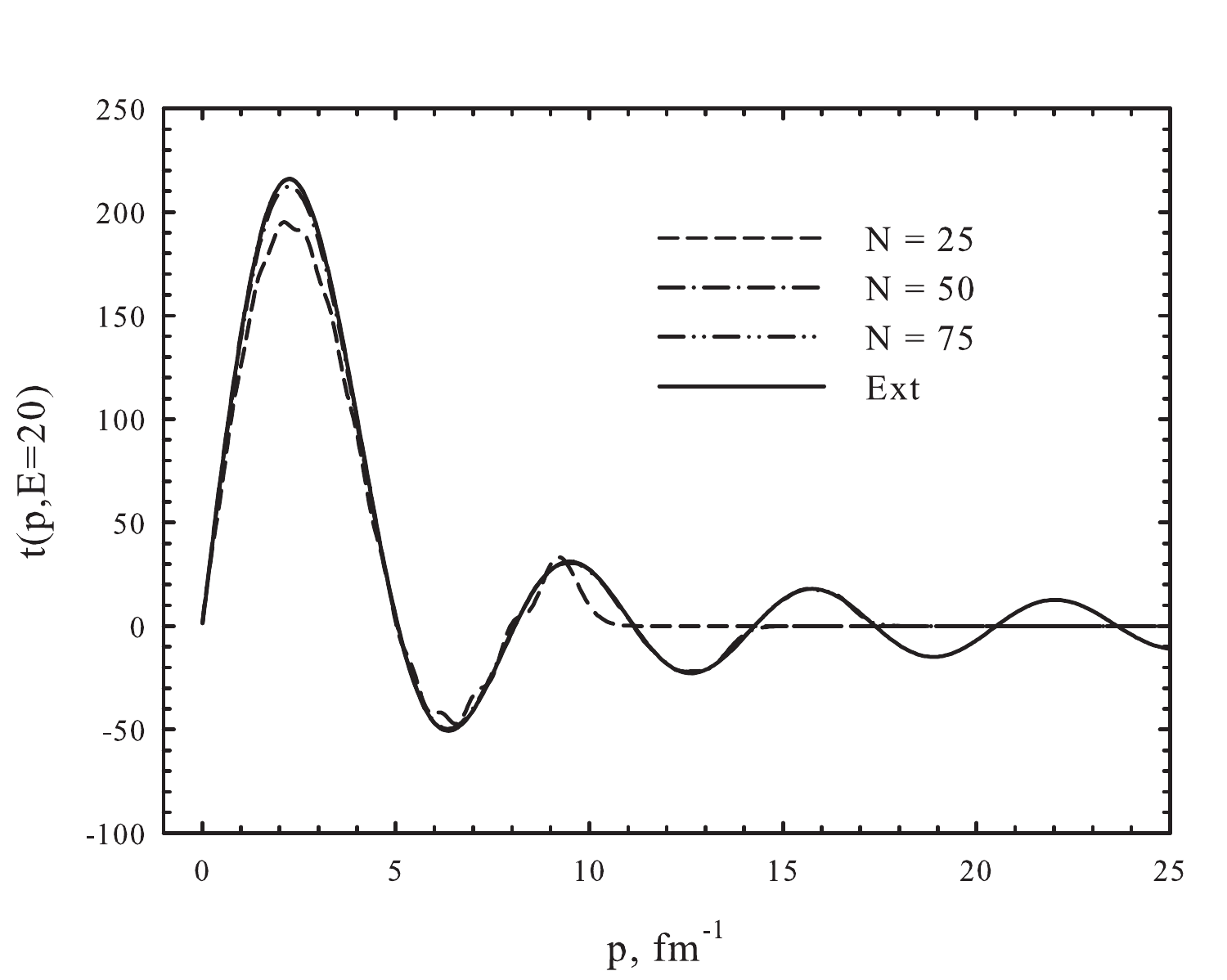}
\vskip-3mm\caption{ Convergence of the T-matrix for the square-well
potential.\,\,Results are obtained for $E=20.0$ MeV and with the
oscillator length $b= 1$ fm }\label{Fig:TmatrSW_vsN}
\end{figure}

Contrary to the Gaussian and square-well potentials, the T-matrix
expansion coefficients for the exponential and Yukawa potentials are
more strongly dependent on the energy.\,\,This is explicitly
demonstrated in Fig.\,\,\ref{Fig:Tmatr_SW_EvsbN}, where the
expansion coefficients for the T-matrix are displayed for a fixed
oscillator length $b$ and for five different values of energy $E$.

The T-matrix as a function of the momentum $p$ for the square-well
potential is shown in Fig.\,\,\ref{Fig:Tmatrix_SW_MS}.\,\,These
results are obtained with $N=300$ oscillator functions.\,\,In Fig.
\ref{Fig:Tmatrix_SW_MS}, we compare the calculated T-matrix with the
exact one (marked as Ext) represented by
Eq.\,\,(\ref{eq:205}).\,\,One can see that the calculated T-matrix
coincides with the exact one in a wide range of momenta $p$.\,\,The
upper limit of the range depends on the number of functions $N$ and
the oscillator length $b$ and, in view of the properties of
oscillator functions, can be expressed as
\begin{equation}\label{eq:210}
p\leq\frac{\sqrt{4N+2L+3}}{b}.
\end{equation}
Thus, the smaller the oscillator length, the larger is the range of
momenta, which can be covered by $N$ oscillator
functions.\,\,Indeed, with $b=2.0$ fm, the nonzero values of the
T-matrix are obtained for $p \leq$ $\leq 17$~fm$^{-1}$, which is in
accordance with formula~(\ref{eq:210}).

So far, to be on a safe side, we used the large basis of functions
($N=300$) to describe the bound and continuous-spectrum
states.\,\,Now, we are going to determine a minimal set of
oscillator functions, which gives solutions with necessary
precision.\,\,We again turn our attention to the square-well
potential, because the wave function and the T-matrix for this
potential are obtained in a simple analytic form. This helps us to
verify the precision of our calculations.\,\,The exact
(\ref{eq:205}) and calculated T-matrices determined with $N=25,$ 50,
and 75 oscillator functions are shown in
Fig.\,\,\ref{Fig:TmatrSW_vsN}.\,\,As is seen, 25 functions cannot
provide with a good precision for the T-mat\-rix.\,\,The calculated
T-matrix noticeably deviates from the exact one, whereas the
T-matrix calculated with $N=50$ and $N=75$ is sufficiently close to
the exact T-matrix.\,\,We note that the larger the number of
oscillator functions involved in calculations, the larger is the
range of momenta $p$, where the T-matrix is described by these
functions.\,\,Note that this number of functions is consistent with
the results presented in Figs.\,\,\ref{Fig:GRState_G&E} and
\ref{Fig:PhaseConv_G&E} as for the convergence of calculations of
the bound state and the phase shift.

A similar picture is observed for the Gaussian and exponential
potentials.\,\,Unfortunately, we do not know the exact T-matrix for
both of them.\,\,For these potentials, the T-matrix calculated with
$N=300$ oscillator functions can be considered as ``exact'', as this
number of functions provides us with a stable solution and the exact
phase shift.\,\,With this definition, the calculations with $N=50$
and $N=75$ basis functions are almost indistinguishable for the
``exact'' T-matrix.\,\,As for the Yukawa potential, one needs at
least $N=125$ to be close to the ``exact'' T-matrix.

\subsection{Asymptotics}

In this section, we consider the asymptotic behavior of the T-matrix
as a function of $n$ for large values of $n$.\,\,In Refs.
\cite{kn:Fil_Okhr, Filippov:1980mn}, it was shown that the expansion
coefficients for wave functions have the following asymptotic form:
\begin{equation}\label{eq:320}
C_{n}=\left\langle n|\Psi_{kL}\right\rangle \simeq\sqrt{2R_{n}}\Psi
_{kL}\left(  bR_{n}\right)\!.
\end{equation}
Thus, the expansion coefficients $t_{n}=t_{nL}(b,k)$ for the
T-matrix can be represented as
\[t_{n}=t_{nL}\left(  b,k\right)  =\left\langle\! n\left\vert \widehat
{V}\right\vert \Psi_{kL}\!\right\rangle\simeq\]\vspace*{-7mm}
\begin{equation}\label{eq:321}
\simeq\sqrt{2R_{n}}V\left(  bR_{n}\right)  \Psi_{kL}\left(  bR_{n}\right)
=V\left(  bR_{n}\right)  C_{n},
\end{equation}
where
\begin{equation}\label{eq:322}
R_{n}=\sqrt{4n+2L+3}
\end{equation}
is the turning point for a classical harmonic oscillator in the three-dimensional space.

In Ref.\,\,\cite{kn:VA_PR} (see also \cite{2004JPhA...37.7769B}), it
was discovered that there is another contribution (which was called
as the short-range (SR) contribution, while the asymptotic term in
Eqs.\,\,(\ref{eq:320}) and (\ref{eq:321}) is called as the
long-range (LR) contribution) to the asymptotic form, which relates
the expansion coefficients $C_{n}$ and $t_{n}$ with the wave
function and the T-matrix
\begin{equation}\label{eq:323}
\left\langle n|\Psi_{kL}\right\rangle \simeq\left(  -1\right)  ^{n}
\sqrt{2R_{n}}\Psi_{kL}\left(  R_{n}/b\right)\!,
\end{equation}\vspace*{-7mm}
\begin{equation}\label{eq:324}
t_{nL}  =\left\langle n\left\vert \widehat{V}\right\vert \Psi
_{kL}\right\rangle \simeq\left(  -1\right)
^{n}\sqrt{2R_{n}}t_{L}\left( R_{n}/b,k\right)\!.
\end{equation}
Equations (\ref{eq:320}) and (\ref{eq:321}) establish some relations between
the expansion coefficients $C_{n}$ and $t_{n}$ and the wave function and the T-matrix
for the large values of momentum $p=R_{n}/b$.

Let us consider the calculation of $t_{n}$ for the square-well
potential in more details.\,\,To determine $t_{n},$ one has to
calculate the integral
\[
t_{n}=t_{n}\left(  b,k\right)
=V_{0}\mathcal{A}_{kL}\sqrt{\frac{2}{\pi}
}k\!\!\int\limits_{0}^{a}\!\!drr^{2}\Phi_{nL}\!\left(  r,b\right)
j_{L}\!\left( k_{0}r\right)=
\]\vspace*{-7mm}
\[
=V_{0}\mathcal{A}_{kL}\sqrt{\frac{2}{\pi}}kb^{3}\!\!\int\limits_{0}^{a/b}\!\!d\rho\rho
^{2}\Phi_{nL}\left(  \rho,1\right)  j_{L}\left( bk_{0}\rho\right)\!,
\]
where $\rho=r/b$.\,\,If $a/b\gg1$ or $a/b>\sqrt{4n+2l+3}$, then the
integral
 can be extended to infinity, and it gives the expansion coefficients for
Bessel functions.\,\,Thus, we obtain the long-range
approximation.\,\,For small values of the ratio $a/b\ll 1$ or
$a/b\ll$ $\ll\sqrt{4n+2l+3}$, we can use the approximate formula for
oscillator functions (Laguerre polynomials).

It was also demonstrated in \cite{kn:VA_PR} that, in some cases
(which depend on the value of oscillator length $b$ and the shape of
a potential), the SR contribution is much larger than the LR
one.\,\,In addition, one has to consider both of them in some cases.
Unfortunately, we do not know the exact asymptotic form of the wave
function $\Psi_{kL}(p)$ in the momentum space and the T-matrix for
large values of $p$ as well.\,\,It makes difficult to realize the
short-range approximation.

Note that the knowledge of the asymptotics for the wave function and
the T-matrix allowed us to formulate strategies (see
Ref.\,\,\cite{kn:AM_AJP} and also Ref.\,\,\cite{J-matrix2008.117})
to obtain convergent results for the phase shift with a minimal set
of oscillator functions.

We have explicit forms of the wave function and the T-matrix only
for the square-well potential.\,\,We will use it to check the
asymptotic behavior of the T-matrix.

\begin{figure}%
\vskip1mm
\includegraphics[width=\column]{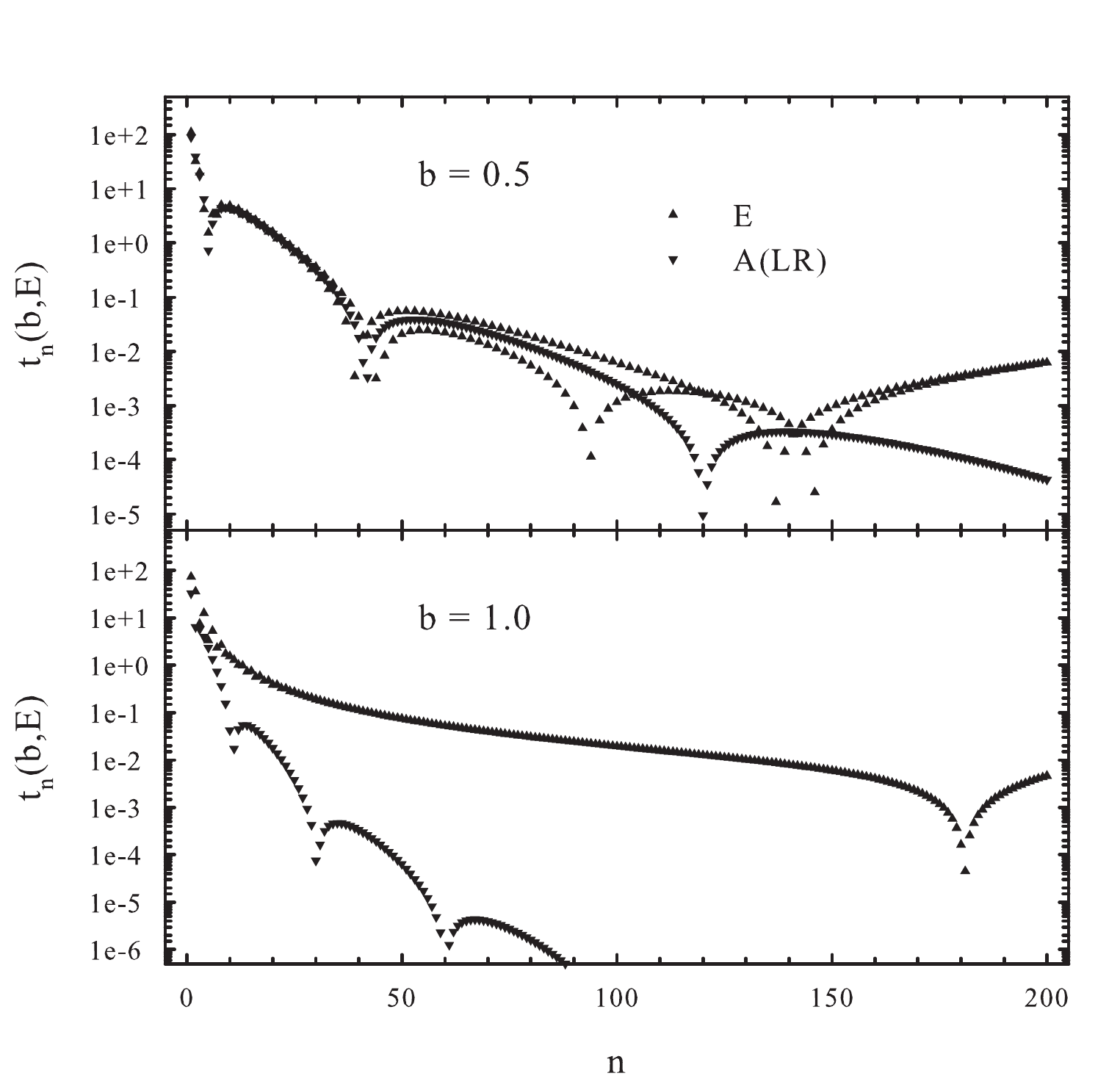}
\vskip-3mm\caption{ Exact (E) and asymptotic (A(LR)) forms of
expansion coefficients $t_{n}$ calculated with the exponential
potential for the energy $E=10.0$ }\label{Fig:Tmatr_Exp_E&A}
\end{figure}

\begin{figure}%
\vskip1mm
\includegraphics[width=\column]{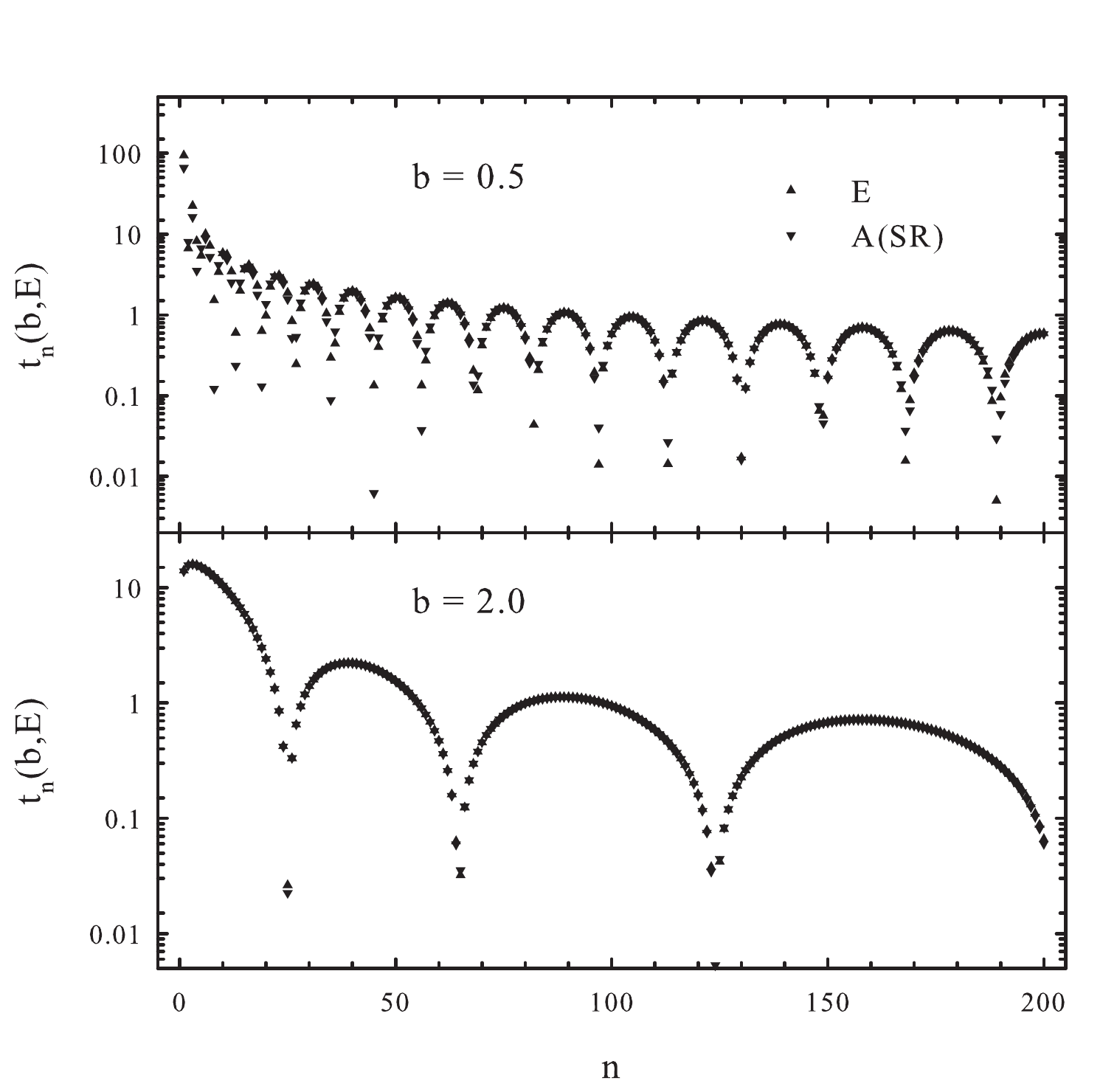}
\vskip-3mm\caption{ Expansion coefficient for the exact T-mat\-rix
(E) compared with the asymptotic short-range form (A(SR)).
Calculations are made for $E=10.0$ with the square-well potential
}\label{Fig:Tmatr_SQW_E&A}
\end{figure}

In Fig.\,\,\ref{Fig:Tmatr_Exp_E&A}, we compare the calculated (E)
and asymptotic long-range (A(LR)) forms of the T-matrix with the
exponential potential.\,\,As is seen, the asymptotic long-range form
is valid for small values of $b$.\,\,Starting from $b=1.0$, the
asymptotic form is much smaller than the exact form.\,\,A similar
picture is observed for the Gaussian potential.\,\,However, for the
Yukawa and square-well potentials, the long-range form gives a very
small contribution comparing with the exact form.

In Fig.\,\,\ref{Fig:Tmatr_SQW_E&A}, we demonstrate how the
asymptotic short-range approximation works for the square-well
potential.\,\,We show results for two values of oscillator length
$b$: the smallest ($b=0.5$)  and largest ($b=2.0$) ones.\,\,As is
seen, the asymptotic short-range form is valid for small values of
$b$.\,\,For large values of $b,$ the asymptotic form coincides with
the exact one for the whole range of the quantum number $n$.

It should be stressed that the long-range asymptotic form gives zero contribution
to the presented results due to a specific shape of the square-well potential.

\section{Conclusion}

We have studied the properties of the T-matrix in the discrete
oscillator representation.\,\,It is demonstrated that the T-matrix
in the oscillator representation can be presented in the vector and
matrix forms.\,\,The vector form is suitable for investigating the
T-matrix on the half-on-shell space, while the matrix form is more
appropriate for the full off-shell space.\,\,The set of linear
equations for the T-matrix expansion coefficients is deduced for
both vector and matrix forms.

We calculated the T-matrix expansion coefficients for four different
potentials.\,\,It is shown that the rate of decreasing of the
T-matrix expansion coefficients depends on the shape of a potential
and on the oscillator length.\,\,It is also shown that the T-matrix
expansion coefficients slightly depend on the continuous-spectrum
state energy.\,\,We recall that the energy of scattering states is
considered in the range $0\leq E\leq 20$, which is typical of the
low-energy nuclear processes.

It is shown that the calculations of the T-matrix in the discrete representation is a
reliable way for obtaining the information on the behavior of a quantum mechanical system.

We have investigated thoroughly the asymptotic properties of the T-matrix in
the discrete space and established a relation with its asymptotic form in the continuous
coordinate and momentum spaces.

The present discretization method can be easily extended for the T-matrix of
real physical systems such as many-channel and many-cluster systems.

\vspace*{-5mm}
\rezume{%
В.С.\,Василевський, М.Д.\,Солоха-Климчак} {Т-МАТРИЦЯ В
ДИСКРЕТНОМУ,\\
ОСЦИЛЯТОРНОМУ ПРЕДСТАВЛЕННІ} {Досліджено властивості Т-матриці для
зв'язаних станів та станів неперервного спектра у дискретному,
осциляторному представленні. Дослідження проводяться для модельної
проблеми~-- частинка в полі центрального потенціалу. Виведено
систему лінійних рівнянь, розв'язок яких визначає коефіцієнти
розкладу Т-матриці по осциляторних функціях. Ми вибрали чотири
потенціали (гаусівський, юкавівський, експоненціальний та потенціал
прямокутної ями) для демонстрації особливостей Т-матриці та її
залежності від форми потенціалу. Ми також вивчаємо як коефіцієнти
розкладу Т-матриці залежать від параметрів осциляторного базису~--
осциляторної довжини та числа базисних функцій, залучених у
розрахунках.}


\end{document}